# Breaching the Future: Understanding Human Challenges of Autonomous Systems for the Home

**Tommy Nilsson, Andy Crabtree, Joel Fischer and Boriana Koleva**
School of Computer Science
University of Nottingham
**{firstname.lastname}@nottingham.ac.uk**

## ABSTRACT

The domestic environment is a key area for the design and deployment of autonomous systems. Yet research indicates their adoption is already being hampered by a variety of critical issues including trust, privacy and security. This paper explores how potential users relate to the concept of autonomous systems in the home and elaborates further points of friction. It makes two contributions. One methodological, focusing on the use of provocative utopian and dystopian scenarios of future autonomous systems in the home. These are used to drive an innovative workshop-based approach to breaching experiments, which surfaces the usually tacit and unspoken background expectancies implicated in the organisation of everyday life that have a powerful impact on the acceptability of future and emerging technologies. The other contribution is substantive, produced through participants efforts to repair the incongruity or 'reality disjuncture' created by utopian and dystopian visions, and highlights the need to build social as well as computational accountability into autonomous systems, and to enable coordination and control.

## 1. INTRODUCTION

Progress in the field of artificial intelligence, machine learning and ubiquitous computing is paving the way for a range of domestic systems capable of taking actions autonomously, largely based on input from sensor-based devices. Systems such as the NEST smart thermostat [1] which predicts occupancy to regulate temperature, or services built around the Amazon Dash Replenishment API [2], are slowly finding their way into the everyday life. Such systems are said to enable a more efficient, convenient and healthy lifestyle [3] and spark market optimism, with analysts suggesting that the global value of the smart home

**Corresponding author:**
Tommy Nilsson (Orcid ID 0000-0002-8568-0062)
Email: tommy.nilsson@nottingham.ac.uk
Tel. 0115 95 14244

Andy Crabtree (Orcid ID 0000-0001-5553-6767)
Joel Fischer (Orcid ID 0000-0001-8878-2454)
Boriana Koleva (Orcid ID 0000-0003-4605-424)

sector is set to exceed £100 billion by 2022 [4].

However, it is important to treat such predictions with a degree of caution as they are based on extrapolations from the current rate of technical progress that fail to take into account potential non-technical pitfalls that might impact mainstream adoption. Novel technology has the habit of introducing novel human problems, and domestic autonomous systems are no exception. Yang and Newman [5] found, for instance, that users of the NEST thermostat have frequent difficulties in understanding its behaviour. Similarly, Rodden et al. [6] found that the prospective users of autonomous domestic systems were concerned about the loss of control and level of personal data harvesting enabled through widespread sensing.

Studies such as these suggest that social expectations regarding the intelligibility and trustworthiness of autonomous systems are as important to their uptake and use as any technological benefits. Yet despite these findings, people's expectations towards domestic autonomous systems remains a relatively poorly understood domain [7]. Following Rohracher's [8] call to engage prospective users from the very envisioning stage of development, we have developed a novel workshop-based methodology that exploits utopian and dystopian scenarios as tools enabling sociologist Harold Garfinkel's "incongruity procedure" [9] or "breaching experiments" [10] as this procedure is more commonly known. The purpose of the procedure is to provoke (as in call forth and make visible) the taken for granted background expectancies that organise familiar settings of everyday life, which in turn impact the adoption and use of new technologies within those settings.

Accordingly we created and explored with 32 potential end-users in focus group workshops a set of provocative contra-vision [11] scenarios to probe and elicit social expectations that impact the uptake of autonomous systems in the home. Our analysis of workshop participants' responses to the scenarios reveals background expectancies centring on *computational accountability* and the legibility of autonomous system behaviour; *social accountability* and the compliance of autonomous behaviours with social norms; *coordination* and the need to build the human into the behaviour of autonomous systems; and *control*. Each of these key topics is formed by and brings with it a range of ancillary expectations that impact the adoption of autonomous systems in the home, and open up design possibilities to enable developers to gear autonomous domestic systems in with the non-technical expectations that govern their uptake in everyday life.

## 2. APPROACH

Our attempt to elicit background expectancies impacting the adoption of autonomous systems in the home merges *breaching experiments* [10] with *scenario-based design* [12] and *contra-vision* [11] to create provocative visions of the home of the future that intentionally disturb common sense reasoning and create incongruities or 'reality disjunctures' whose repair surfaces taken for granted background expectancies that impact the uptake of future technologies in everyday life. Below we describe each of these three methodological components and their role in engaging prospective users with the design of future and emerging technologies.

### 2.1 Breaching Experiments

The notion of breaching experiments has previously been employed in design to understand 'in the wild' deployments of technology [13, 14, 15, 16]. The approach derives from Harold Garfinkel's ethnomethodological brand of sociology [17]. Much like Garfinkel's portrayal of the sociologist, designers commonly select "familiar settings" such as familial households or workplaces, to make prospective technological (rather than social) systems accountable and to motivate particular undertakings. Take Mark Weiser's much cited Sal scenarios [18, by way of example.

As Garfinkel [17] observes, no matter what the sociologist (or the technologist) might have to say about familiar settings, it is also the case that the members of society have their own common sense understandings of familiar settings and employ them as a scheme of interpretation enabling concerted action and the ongoing conduct of social life. Common sense understandings furnish members with background expectancies drawn upon in situ, as contingencies dictate, to both recognise events as events-in-familiar-settings and to enforce compliance with expectations of action in familiar settings.

The latter point is of particular consequence to design as it speaks to the moral ordering of everyday life. It is not then that background expectancies are only used to interpret and make sense of events in familiar settings, but to assess and *enforce* their social acceptability too. We take it that this applies as much to projected technological events that implicate people located in familiar settings as it does to actual events here and now, and that design may therefore be usefully informed through an investigation of the background expectancies that members know and use to understand and order familiar settings of everyday life.

Indeed, given the practical indispensability and moral importance of background expectancies, we might expect members to have much to say about them. Curiously, however, Garfinkel observes that the member is typically at a loss to tell us specifically what the expectancies consist on any occasion of inquiry. Indeed, "when we ask him about them he has little or nothing to say." So, what to do?

In his own efforts to get to grips with the orderliness of everyday life, Garfinkel tells us that for background expectancies to come into view one must either be a stranger to its "life as usual" character or become estranged from it. Given that he wasn't a stranger, he developed a procedure of estrangement, which he called the "incongruity procedure" [9] or "breaching experiments" [10, 17]. While these are often associated with making trouble, as Garfinkel characterised them that way, proponents of the approach in design have demonstrated that trouble is not essential or necessary to their conduct [14]. Rather, and to borrow from Garfinkel, we might say that seen from the members' perspective, compliance with background expectancies turns upon a person's grasp on what are commonly seen and understood to be the "natural facts of life in society." It follows from this that the firmer a person's grasp on the natural facts of life in society – and we assume that a firm grasp of such facts is key to one's social competence – then the more severe should be their "disturbance" when those facts are "impugned as a depiction of his (or her) real circumstances."

In more prosaic terms, we take it that "depictions" of real circumstances – of which there may be many forms including storyboards, scenarios, design fictions, lo-fidelity prototypes, etc. – which breach the "natural facts of life in society" may throw the background expectancies that order everyday life into relief. These depictions need not necessarily make trouble. They need only "disturb" common sense understandings of everyday life. It may then be possible, as it became possible for Garfinkel, to "call forth" or make visible the taken for granted background expectancies that order everyday life. As detailed in the following sections, we thus seek to exploit established design methods – scenarios and contra-vision in this case – to intentionally disturb common sense reasoning and surface the usually tacit and unspoken background expectancies that order domestic life and impact autonomous systems for the home.

### 2.2 Scenario-Based Design (SBD)

SBD [19] is a well-established approach towards user-centred design, which emerged in HCI during the 1990s. Scenarios are stories about people and their activities. They help designers focus attention on people and their tasks, which are often left implicit in technological renderings (e.g., software specifications) of systems and their application. Scenarios can be elaborated on paper, slideshows, video, storyboards, etc. They are construed of as "soft" prototypes that provide "minimal context" exposing not only the functionality of a proposed system but specific claims about the user experience [20], which may be assessed by potential end-users.

In approaching the development of scenarios for autonomous domestic systems we found ourselves, like

Rodden et al. [6] before us, confronted by a technology that has "yet to be realised". However, this is not to say that we were starting with a blank slate. As Reeves [21] highlights in examining the origins of ubiquitous computing, technological projections or envisionments are grounded in a "milieu of existing and developing socio-technical infrastructures and innovations, drawing upon developments in diverse technologies."

Accordingly, the initial stages of scenario development involved reflecting on research on autonomous systems and technologies that might impact the home, some of it based on research we were involved in [e.g., 22], some based on research being done by others [e.g., 5], mixed with a degree of speculation on our part about where such developments and technologies could be heading. Initial topics thus included microgeneration and the smart grid [e.g., 23], automated scheduling of energy infrastructure use [e.g., 6], automated scheduling of appliance use [e.g., 24], and appliance diagnostics [e.g., 25], all of which were framed by what we perceived as general concern in the literature with sustainability and the promise that the smart and indeed autonomous home of the future could deliver on this in manifold respects [e.g., 26].

These initial ideas were subsequently developed into *textual* outlines for future scenarios. We developed two outlines per scenario, one focusing on full automation, the other on human in the loop following Yang and Newman's [5] emphasis on the need to balance system autonomy with user engagement, e.g.,

**Optimising for sustainability - full automation**

Donald has just moved into a new smart home. He has never bothered about the ecology and believes that all the talk on global warming and environmentalism is nothing but superstition. One morning when leaving his house to go to work, he's horrified to find that his car has disappeared overnight. He rushes back into the house frantically searching for his phone to call the police. Suddenly he hears a voice coming from the speakers of his smart home IoT system. The voice announces that Donald's car has been found to be producing an unacceptably high level of $CO_2$ emissions. To prevent further damage to the environment, the IoT system has sent the car to a local recycling company. As a replacement, the system has ordered a bike, which is waiting for Donald in his garage. He is told that not only is a bike far more environmentally friendly, but riding a bike will also help him address his increasingly high cholesterol levels.

**Optimising for sustainability - user engagement**

One morning after waking up, Donald's smart phone starts reading out a voice message informing him that his car is producing a dangerously high level of $CO_2$ emissions. He is recommended to have his car recycled and get a new one. The application offers him a list of nearby recycling companies along with contact details. Donald laughs out at the message and decides to ignore it. While preparing a breakfast, his smartphone delivers another message. This time it urges him to do something about his high cholesterol levels, stating that his food is not particularly healthy. While getting ready to leave his house, Donald lights a cigarette. This triggers another voice message and sends it echoing across the house. It explains that smoking is not only unhealthy, but also pollutes the environment. Angrily, Donald slams the doors shut and departs to his work. Donald endures multiple similar messages throughout the day. Nonetheless, in the long run he finds it difficult to ignore all of these recommendations. His lifestyle is gradually and subtly adjusted towards a greater sustainability.

The scenarios included microgenerated energy production and use, optimising energy consumption, home automation (including turning lights, heating and security systems on and off), diagnosing faulty domestic appliances and, as above, optimising for sustainability. In the course of reviewing them it became evident that the *absurdities* [27] they contained might be something we could leverage more productively if we adopted a contra-vision approach to render provocative utopian and dystopian visions instead of contrasting full automation versus human in the loop.

The scenarios were subsequently revised, with iteration involving other members of our research lab who were engaged in smart home research and IoT device construction to ensure the feasibility of what we were proposing. While centred around hypothetical domestic situations, we nevertheless felt it important that the scenarios not drift too far beyond what could reasonably be expected from such systems if our findings were to be of any practical use. The result of the revisioning process was that we ended up with *visual* scenarios (e.g., Fig.1) focusing on four partially overlapping areas where autonomous systems are expected to make a substantial impact in the home:

– **Temperature regulation:** The use of wireless sensor networks for optimisation of indoor temperature is a well-documented area in ubiquitous computing [28] and one that is well within the reach of current technological capabilities. Studies of emerging solutions have demonstrated the technical viability of regulating indoor temperature automatically by monitoring environmental variables, such as occupancy [29]. Yet research has likewise found that such autonomous systems run the risk of rendering their operation unintelligible to users, which in turn can result in confusion and impede a sense of trust [1]. Our scenarios feature an autonomous system that not only controls heating but also ventilation and air conditioning. It explores the tension between autonomous regulation driven by sustainability and its impact on inhabitants.

– **Auto purchases:** Various auto-purchasing and reordering mechanisms represent another autonomous technology with a potentially strong impact on future smart homes. Semi-automatic solutions, such as the Amazon Dash button are already showing promising results [2] and studies suggest fully AI controlled restocking solutions in the future [4]. If coupled with

auto-delivery systems, these technologies

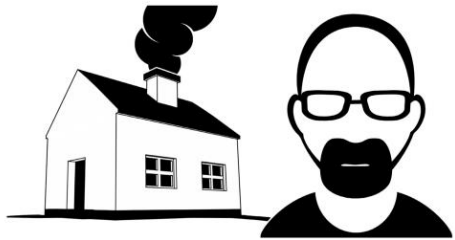

"Donald" - a not too technically proficient smart home resident struggling to make sense of the technology in his surrounding.

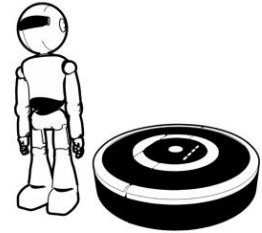

A range of solutions - ranging from sensors all the way to roombas and other robots - manages the household's daily operation.

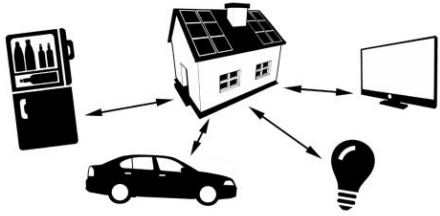

The spectrum of daily life activities managed by the smart home leads to frequent mismatches between system decisions and Donald's preferences

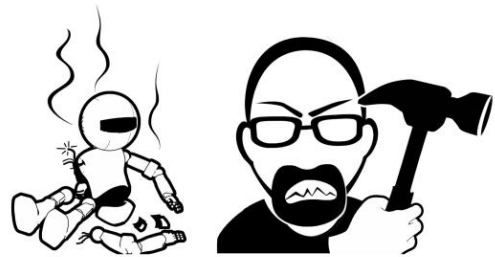

Failing to grasp the cause behind inappropriate system behaviour, Donald often takes out his frustration on his smart home devices.

**Figure 1.** Donald and his autonomous home.

could, for example, be used to prevent essential food items from running out by keeping the fridge stocked. On the other hand, studies have also been flagging that physical sensing of products in the home, computational learning and the prediction of need is problematic, making it difficult for fully autonomous systems to achieve desirable results on a consistent basis [30]. Our scenario explores the tension between auto-reordering and consumer choice and control.

– **Energy regulation:** Many households are already starting to benefit from microgeneration technologies, such as solar panels, which not only produce energy in an environmentally friendly manner, but also help keep the electricity bills low. The fluctuating availability of alternative energy sources, such as solar power, requires continuous monitoring and regulation of consumption. The delegation of such activities to autonomous systems represents a viable alternative that has already been explored in previous studies [6, 31]. Studies have at the same time also been suggesting that the contingencies associated with the everyday life make a seamless operation of fully autonomous solutions problematic [24]. Our scenario explores the tension between autonomous energy regulation in the home and its impact on inhabitants.

– **Lifestyle monitoring:** Sensors are capable of monitoring multiple parameters of our life styles,

ranging from biodata, such as heart rate or sleep patterns, to environmental variables such as humidity and air pollution. This in turn enables a range of solutions promoting a healthy lifestyle by giving the user visual or auditory reminders and clues that are intended to influence their behaviour [32]. Such solutions have been noted by previous research as representing a double-edged sword. Yang et al. for instance have found that such systems might succeed at least temporarily in adjusting user's patterns of behaviour but are prone to causing new problems, such inadvertently providing poor or misleading recommendations [33]. Our scenario explores the tension between autonomous regulation of one's life style and individual autonomy.

### 2.3 Contra-vision

As noted above, we complemented our scenario-based design activities with a contra-vision approach [11]. This conveys two contrasting yet comparable representations of the same technology on the premise that this can help elicit a wider range of reactions from prospective users than what could normally be gained by conveying only one perspective. The point and purpose of these contra-visions is not to enable exploration of implementation problems that might occur as users seek to incorporate new technologies, such as autonomous systems and virtual agents, into their daily lives. Rather, and as Mancini et al. (ibid.) put it, contra-vision represents a particularly valuable method when researchers have reason to believe that new technology is likely to raise subtle personal, cultural and social issues that can **"potentially jeopardise its adoption"**. We thus developed a number of *utopian* and *dystopian* contra-vision scenarios covering the four areas outlined above, which are further elaborated in Table 1 below.

| Utopian vision | | Dystopian vision | |
|---|---|---|---|
| **Scenario 1: Temperature regulation** | | | |
| 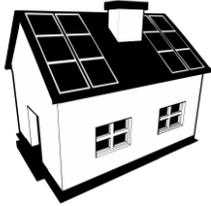 | Temperature in Donald's home is regulated automatically, without the need for any intervention. Everything is taken care of in the background, so that Donald can spend his precious time on more important things. | 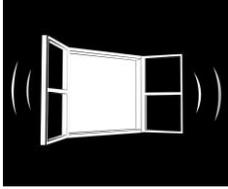 | The windows in Donald's smart home are being opened automatically to ventilate and regulate indoor temperature. The system operates fully autonomously without explicitly notifying its user of the variables that are being taken into account to determine whether to open the window or not. This in turn results in a situation where from the user's point of view the windows are being opened seemingly randomly. |
| **Scenario 2: Auto-purchases** | | | |
| 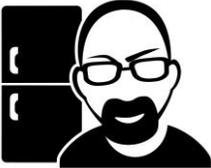 | The smart home is also monitoring food items available in the fridge. Whenever a particular item is running out, the system reorders it to make sure that critical items are always available. | 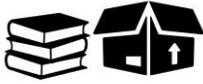 | There are frequent mismatches between what the system determines is needed and what Donald actually wants. Moreover, since the system spends money on his behalf, Donald is slowly losing control over his expenses. |
| **Scenario 3: Energy regulation** | | | |
| 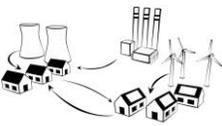 | The domestic system automatically optimises energy consumption in Donald's home in order to keep electricity bills low while contributing to a sustainable society. | 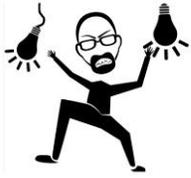 | Electrical devices, such as lights, are being turned off without warning, sometimes even in the middle of being used by Donald. This leads to chaotic situations and even puts Donald in danger. |
| **Scenario 4: Lifestyle monitoring** | | | |
| 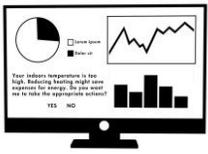 | Through a network of sensors, the autonomous system monitors a range of variables pertaining to the lifestyle of its users. The information serves as a basis for recommendations and even fully proactive actions, such as regulation of air humidity. | 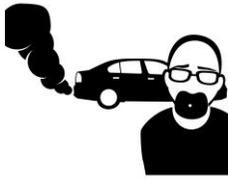 | The system's goal of enabling a healthier environment leads to catastrophic consequences as it proactively decides to send away Donald's old car for recycling and replace it with an electric one in a bid to reduce the household's carbon footprint. |

**Table 1.** Overview of the contra-vision scenarios.

From our own perspective, portraying technology in a purposefully dystopian manner conveys a vision of the future which is essentially broken. Whereas, portraying it in a purposefully utopian manner conveys idealised and idealistic visions of the future. Neither are right, as in correct, and they are not meant to be. Rather, both represent polar extremes (absurd contrasts) which are intentionally designed to disrupt common sense reasoning and surface the unspoken background expectancies that order what

Garfinkel [17] called "an obstinately familiar world" and in turn impact the adoption of future and emerging technologies in everyday life [34].

The combination and repurposing of scenario-based design and contra-vision thus enables a novel approach to the conduct of breaching experiments that provides an early development stage framework for evaluating new technologies at minimal costs while eliciting rich end-user insights. The utopian and dystopian character of the scenarios allows us to provoke and disturb common sense reasoning, which in turn creates a "reality disjuncture" [35] i.e., incongruous experiences of the world that motivate efforts at resolution and repair. It is in repairing the reality disjuncture that the taken for granted background expectancies that people use to make sense of and order familiar settings and scenes in everyday life are "called forth" [17] and thus become available to our consideration.

## 3. PROVOKING REACTION AND REFLECTION

We adopted a focus group, workshop approach to assess the contra-vision scenarios with potential end-users of domestic autonomous systems. We chose this approach because group dynamics can help people to explore and clarify participants' views in ways that would be less easily achieved, if at all, in a one to one interview [36]

### 3.1 Participants

32 people agreed to take part in 2 focus group workshops; 20 were staff members at our place of work (a university) who volunteered on the basis of a free lunch (pizza), and 12 were attendees at the 2016 Mozilla Festival, who volunteered out of interest in the topic. The participants were a mixed bag of designers, IT professionals, enthusiasts, PhD students, researchers, and academics, male and female. As Twidale et al. [37] put it,

> " ... a possible objection to this would be that the small number of end-users might be unrepresentative. However, in one sense, no end-user is unrepresentative in that all end-users' viewpoints and requirements reflect a context in which the system may have to function."

The authors also note that the "limited size" of a study does not stop "major problems" being "discovered and generalised", a matter we will return to in discussion.

### 3.2 Focus group sessions

Each of our focus group sessions ran for approximately 75 minutes and consisted of three key stages:

- **The utopian vision:** We initiated each session by presenting the utopian version of our scenarios. While constituting one of our contra-vision points, this stage also served as an overview of key technology concepts and of how these might ideally be implemented into the home of the future to improve everyday life. Participants were free to interrupt our presentation at any time.

- **The dystopian vision:** Participants were subsequently introduced to the dystopian vision, depicting a broken smart home infrastructure. The aim here was to open up utopian visions to scrutiny and encourage critical discussion and debate. Just as with the previous stage, participants were free to react or raise questions whenever they wanted.

- **Open discussion:** Participants were also encouraged to reflect on and discuss what they saw. We initiated this

stage by asking a broad question, such as *would you be willing to live in a smart home like the one depicted in our scenarios? If not, what would make it work for you?* Participants were given leeway to comment and respond to each other's thoughts and reflections, with occasional interruptions from us to ask additional probing questions around topics raised by participants.

The scenarios were presented via a slideshow, and initially met with enthusiasm. The notion of augmenting domestic life with a range of technologies that operate autonomously was generally well received, and the issues postulated by our dystopian scenarios were frequently waved aside as temporary technical limitations that would eventually be sorted out. It wasn't until participants begun to relate the proposed technologies to their own lives, and to the various social situations that might occur in their homes, that perceived problems started to emerge.

Both focus group sessions were recorded on audio and subsequently transcribed. The transcriptions were used to examine the topical concerns that populated our participant's talk and their reasoning about autonomous systems in the home. Strictly speaking we did not subject the transcripts to thematic analysis in order to "identify patterns within the data" [38] but instead adopted an ethnomethodological orientation to identify the different orders of mundane reasoning exhibited in participants' talk and the endogenous topics elaborated by them (rather than by us as analysts) in the course of their conversations [39]. It is towards explicating the different orders of mundane reasoning and the topical concerns that characterise them that we turn below and elaborate through sections 4 to 8.

## 4. COMPUTATIONAL ACCOUNTABILITY

The importance of designing systems that are accountable to their users has long been recognised in HCI [40, 41]. While the word 'accountable' in ordinary English is often associated with justification, in HCI it generally refers to the legibility or intelligibility of computer systems to end-users [42]. HCI researchers have coined the term "computational accountability" [43] to distinguish this specific kind of accountability. As can be seen in (anonymised) conversational extracts from the focus group sessions below, issues surrounding computational accountability of autonomous systems are not only concerned with accounting for the system's behaviour, but also with what is involved in providing an account, and particularly with making the motives and reasons for the actions of autonomous systems transparent along with the kind of transparency required to provide a sufficient account.

### 4.1 Reasoning about accountability

In considering the idea that autonomous systems could take proactive actions to improve the ecology and sustainability of the household, our participants exhibited a concern in their talk with the legibility of the smart home's *behaviour* to Donald.

**Jack:** It depends how the whole system is being framed for Donald. So if he wants to get a really good deal, if he's been told his house would do things to minimise the amount of energy he uses, he might think "fine", because he wants to keep his bills down. But if he is told "your house will do things because it is trying to be sustainable and environmentally friendly", he might feel like "well, I don't care; I'd rather it do what I want it to do for me."

**Jill:** I wouldn't like to live in a house where I could not account for things that were happening.

This simple interchange between Jack and Jill reflects broad concern amongst our participants with the accountability of autonomous system behaviours. While seeing the potential benefits of autonomous systems, no one wanted to live in an environment where what happened then – the automatic modification of heating levels, the turning on and off of lights, the opening and closing of doors and windows, etc. – could not, as Jill puts it, be *accounted for*. It seems clear then that a critical background expectancy impacting the uptake of autonomous systems is that their behaviours will be accountable to end-users. But what exactly does accountability consist of or require?

### 4.2 Reasoning about motives
A key background expectancy closely bound up with computational accountability became perspicuous when participants were considering the scenario in which the autonomous system in Donald's home decided to replace his car with a newer one. It did this to reduce Donald's carbon footprint without asking him for permission or giving any other prior notification.

**Phil:** But who is it that's telling you that you need to change your car? It could be some kind of a company that you pay to service it and they recommend you to change the car. Is their motive that you're paying for their service? Or is their motive like "let's sell cars based on some dodgy grey statistics that we made up."

**Alan:** People get usually very emotionally attached to their cars.

**Sally:** You might imagine you're in one of those schemes where you get the latest thing, and the latest thing is forced on you, but you have sort of a nostalgic attachment to the previous thing. "This is obsolete now, we'll give you the newest one automatically." You would not like that kind of autonomy.

As Phil makes visible, "motive" matters in making the behaviour of autonomous systems accountable. Of particular relevance and concern is the "telling" of specific motives that drive specific autonomous behaviours and on whose behalf they are acted upon. The "telling" of just what, just why and just who is important and speaks to the active provision of an account as a preface to autonomous behaviour rather than a bolt-on explanation delivered (like Donald's car) after the fact, for as Sally puts it, "You would not like that kind of autonomy". Accountability thus becomes an important part of the conduct of autonomous system behaviours, rather than a narrative that is subsequently appended to action, and speaks more generally of the need to build transparency into the behaviours of autonomous systems.

### 4.3 Reasoning about transparency
The background expectation that transparency should be built into the behaviours of autonomous systems became apparent when participants were confronted with the scenario of temperature regulation in Donald's house, which the system does (in part) by automatically opening and closing windows.

**David:** If you live in a climate where it gets really hot and you need to get cooler air in the evening or morning – if these windows open when I would open them – then great, why not. But why does the slide say the windows are being opened seemingly randomly? That means the system is faulty, the system is not opening them when he would like to open them.

**Sarah:** I guess it kind of works because the windows are opening when it's too hot and closing when it is too cold.

**David:** Who defines when it is too hot? For him? Donald at some point wants the windows open because *for him* it's sometimes too hot or too cold. So if it is seemingly random to him, then the system is not behaving properly.

Like Phil, in raising the matter of motive, David clearly begs the question, on whose behalf is an autonomous system acting? With that goes the expectation that their actions will serve end-user need. But whether acting on Donald or some other unidentified actor's part, David also makes it perspicuous that should an autonomous system's actions appear "seemingly random" it would *not* be seen to be "behaving properly", where "properly" clearly means its actions are intelligible rather than discernibly "random" and something that have to be guessed at as Sarah has to do.

So not only is it expected that an autonomous system operate in tune with the preferences of its users, it is also expected that the grounds upon which it operates on any occasion are visible to end-users. It might otherwise be said that the telling or provision of an account extends to incorporate both the *reasons* for action as well as the motive, which is to recognise that motive alone does not adequately account for action: motive may account for what occasions action (e.g., temperature regulation), whereas reason also accounts for particular actions done with respect to motive (e.g., opening windows to cool the environment). Background expectancies regarding transparency thus extend the accountability requirement to include the 'in vivo' accomplishment of autonomous behaviour, i.e., the *doing* of autonomous behaviour and provision of reasons that account for what is being done.

### 4.4 Reasoning about the nature of transparency
A further background expectancy became apparent in participants' reflection on the previous temperature regulation scenario, which concerns the kind of transparency considered necessary to understanding the behaviour of autonomous systems.

**Researcher:** Would you like to live in a house like this?

**Megan:** Not with Donald.

(General laughter)

**Harry:** The question is whether you would be willing to live in a house where you cannot fathom the things going on around you. Where the house is behaving in ways you cannot understand.

**Megan:** But that happens all the time. My house is too cold and I don't know why.

**John:** Well, we all live in houses we don't understand, unless you know what's going on in a gas boiler.
**Harry:** But you do understand the thermostat, because it is really simple.

It would be easy to think that in expecting the reasons for an autonomous system's behaviour to be made transparent, end-users are also expecting designers to open up a black box of complexity. The above extract disabuses us of this notion. Just as most of us don't understand "what's going on in a gas boiler", and wouldn't understand even if the underlying reasons were carefully spelt out to us, then so it will be with autonomous systems: end-users don't expect to understand the detailed reasoning implicated in their inner workings, and they don't expect to have that level of insight because they have no practical need of it.

A rather more "simple" level of transparency is instead required for the practical purposes of end-users, of the kind provided by a thermostat, for example. Other examples discussed by our participants included articulating the "agenda" of an autonomous system and the underlying "business model", instead of the intricate technical processes going on under the hood, though as noted above, such matters will need to be manifest in the conduct of autonomous system behaviours and not merely as after the fact appendages.

## 5. SOCIAL ACCOUNTABILITY

While discussion amongst our participants was initially oriented to technical considerations of autonomous system's behaviours and their intelligibility, it became increasingly apparent as the conversation progressed that they were also holding the autonomous systems depicted in our scenarios accountable to the broader social context within which they were prospectively situated. *Social accountability* extends beyond expectations concerning the legibility of autonomous behaviour to expectations concerning the appropriateness of autonomous behaviour within a social setting.

### 5.1 Reasoning about acceptability

Concern with social accountability as distinct to computational accountability became evident in participant's consideration of the automatic temperature regulation scenario:

**Rachel:** You wouldn't like the windows opening when you're not at home, regardless how hot it is.
**Megan:** Depends what kind of window it is; not with that kind of window (points to the large open window in the presentation slide).
**Ben:** If you live close by the road, with cars going by while you try to listen to something …
**Adam:** There are also these big bees outdoors …

The extract makes it perspicuous that understanding why an autonomous system is behaving in the way it is (e.g., why it's opening windows) does not automatically mean that end-users will find this behaviour acceptable. On the contrary, people evidently entertain background expectancies regarding the appropriateness of a system's behaviour. Opening windows "when you're not at home", or "while you try to listen to something" or with those "big bees outdoors" all point to the need for autonomous systems to not just be intelligible in their own right, but also in relation to the specific social context within which they operate. Their behaviours, in other words, need to be *intelligibly appropriate* given the specific social conditions in which their actions are embedded 'here and now' (no one is at home, people are busy, there are hazards outside, etc.).

### 5.2 Reasoning about agency and entitlement

Background expectancies bound up with the appropriateness of system behaviour also turn upon considerations of agency and entitlement, as became perspicuous when participants were confronted with the notion that people could order products simply by talking to the autonomous home through a voice interface:

**Sam:** I can imagine with the voice thing, if it would actually start buying stuff… that would be really bad!
**Sue:** If you've got your five-year old kids in the house alone they might go like "chocolate and popcorn!" and the voice replies "OK!" and off we go!
**Jill:** It should be able to know who is actually going to end up paying for it, because there are those examples of kids that run up hundreds of pounds on app purchases because their parents left out an iPad.

As the extract makes visible, it is not only the social conditions in which autonomous systems are embedded that impacts the appropriateness of their behaviour, *who* is implicated in those behaviours also counts. Clearly not everyone has the same rights and privileges in the home and our participants held autonomous systems accountable to mundane expectations regarding how everyday life in the home is socially organised. Thus, and for example, an autonomous system should not reorder "chocolate and popcorn" just because it has run out and the kids want more. Rather, any such action should be accountable to the party who is "actually going to end up paying for it". An autonomous system embedded in the fabric of domestic life must effectively become a social agent whose actions comply with the mundane expectations that order domestic life. Put simply, the system needs to act in a manner consistent with that which would be expected of a responsible human agent.

### 5.3 Reasoning about trustworthiness

A key social accountability expectancy concerns the trustworthiness of autonomous systems and whose interests they serve, as become visible when participants were discussing auto-replenishing systems capable of reordering worn out household items:

**Researcher:** So the next problem that Donald encountered was the problem where random products were being delivered to his doorstep automatically. He didn't order anything and everything was paid for without his consent.
**Jill:** So that's what I wanted to mention that we're actually assuming that you trust your house or you trust your personal assistant. We assume that they do the right thing.

**John:** Say your washing machine gets broken and your house has purchased this specific spare part from one specific supplier. But that particular brand is maybe not in your best interest, because you know, it spoils quicker. It's not in your best interest, but it's in the best interest of the guys who made the washing machine, or some other company out there.

As John makes visible, even if the actions of the autonomous home are intelligible and the end-user knows why it has done what it has done (e.g., bought a replacement part because the washer was broken), it is not evident that doing so is in the end-user's interest. On the contrary, it might "be in the best interests of the guys who made the washing machine, or some other company out there". There were, as we have already seen, multiple instances where our participants questioned whether the behaviour of an autonomous system could really be trusted and its actions carried out on the end-user's behalf. Our participants expected that external interests would exert considerable influence on what goes on in the autonomous home, and found this deeply problematic. It is critical then that autonomous systems not only articulate motives and reasons, but that their actions are clearly accountable to end-user interests. But how?

### 5.4 Reasoning about risks
Clearly autonomous systems raise the risk of widespread exploitation from an end-user viewpoint, championing the interests of external parties at the expense of those who live in the autonomous home. Another closely related key area of risk concerns the potential for widespread data harvesting and the risk this poses to privacy.

**Simon:** The thing about AI technology is that all of this stuff is data and it's stored somewhere and it knows all these things about you and potentially that makes you a target for an attack.
**Researcher:** You never know who else is watching?
**John:** Or who else has a relationship with your house.
**Wilma:** If we start encouraging smart homes to have their own server, so all your data goes back to your cloud and that sends a limited amount of data to the people that need it, once that becomes a viable option, stuff like the legal requirements around how you store people's data, and how you make this clear to them, is going to make it so that if a company is asking for detailed data from your life, then that might be something why you choose not to do business with them.
**Martin:** You need an on-off switch and an awareness model so that you can close your data from going out and you're aware of what data is going where. Everything is happening, I mean the Internet of Things, things that are in your home sending data of yours without you realising it. Or providers giving you a service without actually explaining their business model is another problem.

In considering the implications of our scenarios, our participants' discussion reflected widespread social concern about the impact of intelligent technologies on everyday life and "who else has a relationship with your house" enabled through current cloud-based autonomous infrastructure. Our participants expect that "what data is going where" be made accountable to them by the autonomous home. They posit various mechanisms enabling this, including local data storage mechanisms that make end-users aware of data flow and which allow them to "limit" the flow of data and

"close" it off, as well as "legal" mechanisms that enable end-user choice. Such mechanisms might in turn make the behaviours of autonomous systems in the home accountable to end-user interests.

## 6. COORDINATION
Clearly our participants had background expectancies regarding what was and is acceptable in their homes and hold autonomous systems accountable to them. Expectations that autonomous systems will behave appropriately in relationship to social conditions, social actors, and end-user interests requires that autonomous systems coordinate their actions with the inhabitants of the home. However, unlike traditional computer applications, many autonomous systems are designed to work on the periphery of user's attention. Instead of simply responding to instructions obtained through manual input from the user, these systems typically have to act on sensed data collected from their surroundings. Nonetheless some kind of interface is required and, while drawing on existing literature, we presented a range of alternatives in the course of exploring our scenarios, including traditional GUI dashboards [44], voice-based interfaces [45], virtually embodied agents [46] and even humanoid robots [47]. While the latter were generally seen as too futuristic, the former were seen as relevant to enabling coordination.

### 6.1 Reasoning about voice control
Given the current turn towards voice-based interfaces (e.g., Amazon's Alexa and Google Home) much of the discussion about coordination inevitably revolved around the potential of voice interfaces, though not unproblematically.

**Researcher:** Another type of interface is the voice-based interface, which allows you to interact with a computer in a very natural way by speaking. It can basically wake you up or contact you whenever, it basically just shouts out recommendations or it could use something like motion sensors to deliver relevant messages whenever you're near a device.
**Tom:** Personally I find voice a little intrusive. I don't even like phone calls. And the prospect of this thing going "Hey Tom, you need to eat carrots!" just while I'm talking or doing something else is not something I'm interested in.
**Sarah:** Imagine if you have visitors in the house, so embarrassing!
**Tom:** It's not what I would want.
**Henry:** It depends if this is about all these things shouting at me or is this a single agent I'm interacting with by voice. I sit down in my living room, and I'm like "hey voice-thing tell me the things I need to know about things", and that's useful.
**Fred:** Let's just paint a picture where you sit with your friends and you're eating some pizza, and it says "Hey, can I chat with you in another room?" And you go there and it says "On recommendations from the NHS [National Health Service] blah, blah, blah. You might want to do this. Do you want to do it?" You say yes or no. So it could be like a communication between you and your house.
**Mike:** I'm just visualising my fridge saying "You guys had an awful lot of beer tonight, maybe ..." you know?
**Claire:** What keeps coming up is that it's a relationship. Like the house is – the house becomes a housemaid. It might have your best interest at heart, but if it's socially awkward, or just does things wrong, brings in a bad mood, then it is a bad housemaid. So I think until you can have some form of natural feeling interface which isn't just going to be insulting you, or telling you to do stuff or muttering at you or treating you like a child then that's gonna be a big issue. Making a house that can do something is easy. Making a house work out whether it should is, I think, a lot more difficult.

As our participants' talk makes clear, just as automatic behaviours (e.g., opening windows) are socially accountable to background expectancies regarding appropriateness then so too are coordinative behaviours. Context awareness is key. As Tom puts it, voice-based interventions would be inappropriate "while I'm talking" or "doing something else". Such interventions are seen as being not only potentially inappropriate but "embarrassing" if the underlying systems is not aware of the social circumstances to hand, e.g., that "you have visitors in the house". This, of course, is not to say that voice-based coordination could not be useful, enabling autonomous systems to be seen as an "agent" or "housemaid" working in your "best interest", but that usefulness needs to be balanced with a sensitivity to social context, as underscored by Claire.

### 6.2 Reasoning about timeliness
An issue that is intimately bound up with social context and coordinating autonomous systems with people is the timeliness of coordinative actions.

**Matthew:** If I just got home and I'm carrying 15 shopping bags, I don't want a thing to say "Matthew, have you bought carrots?" It's a real problem, a system that is going to interrupt me. A real person looking at me carrying 15 shopping bags would not say "Have you bought carrots?" I would be angry!
**John:** Like here, if the system is telling me "Do you need a car?" that is not something you need to be told while you are in the car.
**Henry:** It's fundamental that this is about asking for information at a time when it is, you know, not out of tune, and there is something you can do about it.

Evidently an autonomous system that seeks to coordinate with users will be seen as problematic if its actions lack sensible timing. The ability to "interrupt" at the right time thus becomes a vital component of acceptability. As Henry puts it, it is "fundamental" that system behaviour is "not out of tune" with human behaviour and occurs when there is opportunity to do something about it. Achieving coordination, in this sense, thus becomes a matter of orchestrating or synchronising system behaviours with the user's situation.

### 6.3 Reasoning about situated action
As the following extract makes perspicuous, there is more to synchronizing system actions with the user's situation than being aware of what is happening 'here and now'.

**Researcher:** Would you accept a house that tries to change you and your habits?
**Sarah:** When it comes to healthy lifestyle, like you said, you're not going to do it unless you want to. The house can try, but it should never force you. You should opt in to wanting to change. So you should have thing like "I want to have a more healthy lifestyle, but I don't know how to do that, can you help me house?"
**Researcher:** So you're gravitating towards a recommendation system rather than a totalitarian smart home?
**Sarah:** It definitely has to ask for information first, because if it starts recommending you stuff without knowing your situation – like for example, I could be eating a lot of junk food because I'm working 60 hours a week, and healthy food right now is just not making me happy; bad lifestyle to get me through my long work schedule. And although the smart thing here would probably be for the house to suggest that I don't work so hard, here what it's doing instead is making me more miserable.

Context awareness is not just about being sensitive to social circumstance and timing interruptions to gear in which the current situation. That is to say, context awareness is not simply a momentary concern for people, but something that has a *temporal horizon* – e.g., "I'm working 60 hours a week". Coordination thus extends and turns upon gearing in with such things as a "long work schedule". This will require that autonomous systems are furnished with "information" about the user's situation to enable the effective coordination of system behaviours.

## 7. CONTROL
Control is the final major topic surfaced in our focus group sessions. In spite of dealing with autonomous systems, i.e., solutions designed to carry out tasks on the user's behalf, it became increasingly clear as the conversation unfolded that manual control mechanisms enabling users to customise or even terminate system actions would greatly contribute towards the overall acceptability of autonomous systems in a domestic context.

### 7.1 Reasoning about customisation
The need for autonomous systems to be furnished with information about the user's situation to enable effective coordination raises the issue of *customisation* as a distinct form of control.

**Sam:** I must say you have painted a really stupid autonomous system. You paint autonomous homes as something negative.
**Dave:** It's intentionally provocative. The point is that it is extreme because if we talk about less extreme, the responses would be less interesting.
**Sam:** But it's not realistic.
**Dave:** It is realistic. If you live in social housing, they will put solar panels on your roof without asking you. They can turn off your electricity without much notice. You've got central heating which can shut off if there is not enough demand. There are all these things, OK. If you own your own home and are in full control of it, you're fine, but if you imagine a managed house, which a lot of people live in, then it's not impractical to suggest that some of this might happen. So it's not too far removed from what the reality is.
**Sam:** I can't imagine anybody programming an autonomous system in a house that is this stupid.
**Dave:** Really? How long have you been doing computer science stuff?
(Group laughter)
**Sam:** I have faith in the fact that we can build working systems so there's a level of autonomy, a level of smartness, that's the thing. I think in this realm of setting things and turning lights off and stuff, I think it should not do that unless you want it.
**Researcher:** OK, so another problem that Donald ran into was the energy regulation issue. In the first scenario, the smart home started to shut down all the devices automatically. Whereas now it needs some input from the user. So a dashboard would probably look something like this. It would show a popup message encouraging you to do something and offering to take the appropriate actions on your behalf.
**Jack:** I kind of like the negotiations in the message interface.

**Claire:** But there's an awful lot of complexity in that …

**Sarah:** I'm sorry to interrupt you, because we're saying that all these things would be ongoing if not scheduled for a specific time. Why would it not be possible that he schedules exactly the time he wants all these events? You know, so that - of course there might be emergency so he'll have to change the schedule, but this does not happen every day I guess -- so he could schedule like every day I want this to be done every time right before I go to bed or before I visit someone or do something else.

The dispute between Sam and Dave about how realistic our scenarios of the autonomous home might be ultimately surfaces the expectation that an autonomous system should not do things "unless you want it". This triggered discussion of dashboards as a means of users configuring "appropriate actions". The expectation here is that customisation should extend beyond system messages encouraging users to do things to enable users to "schedule events" themselves. In speaking about scheduling events "right before I go to bed" or "before I visit someone" or "do something else" there is a strong sense in which Sarah speaks of customisation as a means of gearing the behaviours of autonomous systems in with the rhythms and routines that are implicated in the user's everyday life.

### 7.2 Reasoning about direct control

On a final point of note our participants also expected that they would be able to exercise direct control over the behaviour of autonomous systems. This was surfaced at various points throughout their discussions but perhaps none so pointedly as here:

**Paul:** You know there is a lot of ideology in the structure of the problem itself. A lot of things are biased ideologically to start with. I see it as a move towards a consumerist attitude that I don't see very likely to work. They can sell you a car that won't start if you drank too much. You might think it's a very good idea until the day that you have somebody who is bleeding to death and you drank a couple of beers. You want to run to the hospital, but the car doesn't start. The most important thing on a machine is the off button. Of course, you could use it if you drink a couple of pints too much at the pub, maybe, if you're irresponsible. But it's your responsibility, so you must have an off switch.

**John:** How do you turn these systems off basically? I mean I appreciate your point, I do think technology should have an off switch, I'm just wondering, what form does this off switch take? If your magical smart future car has an off switch, you don't want that off switch to occur when you're driving 70 miles an hour down the highway.

**Claire:** It's also about all the smart features, I mean how dumb do you make it. If you had a smart on and off, well now I can exceed 70 miles per hour and I can drive drunk, but I can't use cruise control or I can't use automatic gearbox, because I made it into a totally dumb car. So I think it's definitely a sliding scale of how smart do I want this car to be.

The ability to exercise direct control and shut down an autonomous system is evidently an expectation that people hold, but it is not at all a simplistic on/off expectation. While some circumstances may warrant an "off switch" (e.g., overriding an autonomous system when "you have somebody who is bleeding to death"), others that may put users at significant risk (such as switching off while "driving 70 miles an hour down the highway") do not. The issue of control is not only one of balancing risk with human autonomy, however. Bound up with this is the expectation of granular control, of a "sliding scale" of smartness that can be determined by human beings to enable a better fit between autonomous systems and the particular human circumstances in which they are embedded.

## 8. WHAT'S SEEN IN THE BREACH

Our utopian and dystopian scenarios depicting an autonomous future home have been intentionally created and employed to enable breaching experiments that disturb common sense understandings of domestic life. The disturbance creates reality disjunctures and surface taken for granted background expectancies in participants' efforts to repair them. These background expectancies are ordinarily used by people to understand and order action and interaction in what sociologist Harold Garfinkel [17] called "an obstinately familiar world." That is, a world that members are "demonstrably responsive to" in mundane interaction but are ordinarily "at a loss" to tell us just what the expectancies that lend commonplace scenes their familiar, life-as-usual character consist of (ibid.).

Our scenarios have provided means, motive and opportunity to (gently) prod and provoke, and for members to thereby articulate and reflect on, expectations at work in domestic life that are ordinarily taken for granted and left unsaid. In doing so the participants in our study have elaborated distinct challenges for future technological visions. These include the expectations that the behaviours of autonomous systems will be accountable, that their behaviours are responsive to context and gear in with end-user behaviour, and that people can intervene in their operations. More specifically we can say that our participants expect that autonomous systems will be computationally accountable, socially accountable, and that coordination and control are enabled. Below we consider each of these expectations, and their implications for design, in more detail.

### 8.1 Computational accountability

As noted above, computational accountability refers to the legibility or intelligibility of system behaviours. To borrow from our participants, it will not do for the behaviours of autonomous systems to appear "seemingly random", they must be "accounted for". The expectation that the behaviours of autonomous systems will be accountable has close parallels with Dourish and Button's notion of "translucency" [

[42], which emphasizes the importance of a system's ability to give accounts of its own behaviour in order to render its actions legible and thus better support human-computer interaction.

For Dourish and Button computational accountability is about making the inner workings of the "black box" accountable to users. They provide an example of file copying, replacing a general progress bar that glosses computational behaviour with data buckets and the articulation of flow strategies to elaborate the point. A number of studies [e.g., 30, 48, 49] have subsequently established that revealing more of what goes on under the hood of technological systems is often needed to avoid a range of problems, including trust-related issues, that could otherwise negatively impact the user's overall experience. However, when confronted with hypothetical systems operating more or less autonomously, our participants' expectations about computational accountability operate at a different level. Our participants were not so much interested in making the opaque inner workings of autonomous systems accountable, as they were in making their *observable behaviours* accountable.

Our participants thus expect transparency to be built into autonomous behaviours, where transparency means the grounds of behaviour are visible and available to account. The grounds of behaviour were spoken about in terms of "motive" and "reason", where the former articulates what occasions behaviour and the latter articulates what is done by an autonomous system in response to motive. Previous enquiries into autonomous systems, such as a study by Lim et al. [50], have stressed the importance of computer systems being able to articulate the 'why' behind autonomous actions. Our results expand on these findings and paint a broader picture. Of particular concern to our participants is the articulation of on whose behalf autonomous behaviour is occasioned, and the commensurate expectation that whatever is being done is being done to serve and meet end-user need. Consequently our participants expect "simple" accounts of autonomous behaviour, i.e., accounts that articulate what is being done, why and on whose behalf.

Although our findings shift the focus away from accounts articulating the inner workings of autonomous machines, we do find resonance in our participants' talk with Dourish and Button's notion of accountability [42] in the way in which accountability should be expressed. As Dourish and Button (ibid.) put it,

> " ... what is important ... is not the account itself (the explanation of the system's behaviour) but rather accountability in the way this explanation arises. In particular, the account arises reflexively **in the course of action**, rather than as a commentary upon it ..." (our emphasis)

Our participants speak of autonomous systems "telling" users the motives and reasons for their behaviour as a preface to and/or in vivo feature of that behaviour rather than something that is bolted on after the fact to provide a post hoc explanation. Thus, and as Dourish and Button put it, computational accountability "becomes part and parcel of ordinary interaction with a computer system."

The takeaway for design is not simply that autonomous behaviours should be accountable to users. Rather, in expressing taken for granted background expectancies at work in domestic life, our participants have articulated what an account should consist of and look like. Thus we find that the accountability of autonomous behaviours should not be concerned with the inner workings of autonomous machines, but should instead articulate motives and reasons, detailing what is to be done, why and on whose behalf in particular. Furthermore, the articulation of motives and reasons should occur as preface to and/or in vivo feature of autonomous behaviour, rather than an after the fact explanation. For as our participants succinctly put it, users "would not like that kind of autonomy."

## 8.2 Social accountability

Social accountability moves beyond the expectation that autonomous systems will make their behaviours accountable to end-users to instead address expectations regarding the appropriateness of autonomous behaviour. Simply put, autonomous behaviour may be intelligible to end-users but that does not mean they will find it appropriate or acceptable. Acceptability is of longstanding concern in systems design. The Technology Acceptance Model, or TAM, is often cited as a key approach to determining acceptability, being designed to measure a prototype's perceived "ease of use" and "usefulness" [51]. While such an approach has been said to help maximize commercial success of novel systems [52], the TAM framework has also been a target of criticism for its overly generic nature and limited scope [34].

Alternatively, User Experience or UX models focus on "interface quality" and its impact on the '"experiential component" of system use [53, 54], which leads more generally to a concern with usability in HCI. However, as Kim [55] points out, usability is only one aspect of user acceptance, a point underscored by Lindley et al. [34], who also note that usability studies rarely look beyond prototypical implementations to consider broader challenges of adoption in everyday life. Indeed, as our study demonstrates, domestic autonomous systems introduce a range of social concerns that extend well beyond usability, interface quality, usefulness and ease of use to the 'fit' of machine actions with social expectancies. The need for social accountability introduces a broader lens for considering acceptance, encompassing not just the 'product' itself, but also the social circumstances within which it will be embedded and the implications of this for design.

At first glance it may be thought that social accountability is an external prerogative, something that cannot be built into autonomous systems insofar as it turns on user perceptions of what constitutes appropriate behaviour and is therefore a subjective matter. However, this is not case, a) because appropriateness is an intersubjective (social) matter as clearly articulated in the background expectancies our participants *share*, and b) in articulating those background expectancies it became evident that there is much for design to do in terms of supporting or enabling social accountability.

Thus, from a design perspective, it is clear that our participants expect autonomous systems to be responsive to the particular social circumstances in which their behaviours are embedded. There is need then for autonomous systems to take what people are doing into account, not opening windows for example when human behaviour might be disrupted by such an action, and to tailor autonomous behaviours around the social context in which they operate. In other words, in addition to being autonomous, such systems also need to be *context-aware* if their behaviours are to be seen and treated not only as intelligible but as intelligibly appropriate given the specific social conditions in which their actions are situated.

A key expectation in this regard is that autonomous systems effectively exhibit social competence. There is little to be had but trouble from automatically reordering foodstuffs, for example, if it is done without respect to the social consequences of doing so, and not in general but for the particular cohort that inhabits a particular home. Thus autonomous systems need to act with respect to human agency and entitlement and tailor their behaviour around the differential rights and privileges at work in the home. Autonomous systems will in effect need to become social agents whose actions comply with the mundane expectations governing domestic life if they are to assume a trusted place in within it.

That autonomous systems are trustworthy is a critical expectation [56], which also turns on their demonstrably acting in end-users' interests. This is not only a matter of computational accountability and making autonomous behaviours intelligible to people. It is also and effectively a matter of making it visible that actions done are done for you and not, for example, for the benefit of an external party. Thus in addition to exhibiting social competence, the behaviours of autonomous systems must also be accountable to end-user interests if they are to assume trusted status. Of particular note here is the accountability of data – the oil that lubricates the autonomous machine – and transparency of data flows, coupled with tools to enable end-users to limit them and even close them off if it is deemed the machine is not acting in their interests.

The takeaway for design is that social accountability is distinct from computational accountability: the latter speaks to the intelligibility of autonomous behaviours, the former to their appropriateness and acceptability. Social accountability brings with it the need to build context-awareness and trust into autonomous systems. Context-awareness is needed to enable autonomous systems to respond appropriately to the particular social circumstances in which their behaviours are embedded and trust is an essential condition of their uptake in everyday life. It requires that computational agency exhibit social competence and that autonomous behaviour complies with the differential rights and privileges at work in any particular home (i.e., not generalised rights and privileges but situationally specific rights and privileges). Trust also requires that autonomous behaviours are accountable to end-user interests and turns on the transparency of data flows and ability to control them.

## 8.3 Coordination and Control

We treat coordination and control together here as they may be seen to directly complement one another and span a spectrum of expectations to do with the orchestration of autonomous and human behaviours. Our findings also make it visible that context-awareness is seen as key to coordination by our participants to ensure that autonomous systems do not make inappropriate interventions. However, as Bellotti and Edwards [41] point out, in order to become acceptable, systems cannot rely purely on context awareness to do things automatically on our behalf, but should rather involve active input from users at least on some level. Similarly, Whitworth [57] argues that computing systems often have a poor understanding of context, which makes it necessary to give users control in order to preserve *their* autonomy. An issue underscored by Yang and Newman [1], who have argued that optimal user experience should be achieved through balancing machine intelligence with direct user control.

Our study unveils a similar sense of scepticism regarding the ability of fully autonomous systems to correctly assess every given situation and act in line with our expectations on a consistent basis. A key expectation at work here concerns the timeliness of autonomous behaviours and the need to synchronise them with the user's situation. Our participants' expectations regarding timeliness are of particular note, including, but moving beyond, a 'here and now' understanding of timeliness (e.g., raising potentially embarrassing matters while I'm entertaining visitors). Of equal concern is the temporal horizon of action and the need for autonomous systems to be sensitive and responsive to temporally distributed patterns of human behaviour (e.g., long work schedules).

Our participants therefore expect that human input will be required, which they speak about in terms of "customisation" and user-driven "scheduling" of events. In this respect it is expected that customisation would allow users to coordinate the behaviours of autonomous systems with occasional and established patterns of human conduct (e.g., long but not permanent work schedules and the reoccurring rhythms and routines of domestic life). This would allow users to configure appropriate actions and help address the thorny problem of how autonomous systems are to develop an awareness of context, including what it is appropriate to do and when it is appropriate to do it. In effect, the problem of learning context is offloaded onto the user to some extent through the provision of coordination mechanisms that enable users to gear the behaviours of autonomous systems in with everyday life in the particular domestic environments they are deployed and used.

The qualification 'to some extent' is important here, for no matter how much they learn about everyday life, and how smart they become, autonomous systems will never be able to anticipate and respond appropriately to *all* social circumstances [58]. Thus it is expected that end-users will be able to exert direct control over autonomous systems. However, the expectation is not that users will simply be able to turn autonomous systems off – that may only be necessary in critical situations – but rather that direct control can be exercised on a "sliding scale". In effect, it is expected that the intelligence built into autonomous systems would be subject to granular control, with levels of intelligence being increased and decreased according to circumstance.

The takeaway for design is that potential end-users do not expect autonomous systems to act independently of user input. End-users expect that autonomous systems will be responsive to context and act in timely fashion, where 'timely' means they are responsive not only to what happens 'here and now' but also what happens over time. It is thus expected that autonomous systems will gear their behaviours in with occasional and established patterns of human conduct, and that mechanisms be provided to enable users to configure autonomous behaviours around human schedules to enable effective orchestration and synchronisation. It is expected too that users will be able to exercise vary levels of control over the behaviour of autonomous systems in order to make them responsive to the inevitable contingencies of everyday life in the home.

### 8.4 Design challenges

The background expectancies articulated by our participants elaborate several distinct design challenges for the development of autonomous systems for domestic use. These include:

−  Building *accountability* into the behaviours of autonomous systems by articulating motives and reasons for autonomous behaviours as a preface to and in vivo feature of those behaviours. What we mean here is different from the literature stressing the need to reveal inner workings of a system (e.g., [42]); instead we are concerned with the overarching motivations and agendas that drive an autonomous system's decision-making.

−  Building *context-awareness* into autonomous behaviours to enable autonomous systems to respond appropriately to the particular social circumstances in which their behaviours are embedded. Drawing on a rich body of existing research in context-aware systems (e.g., [41]), the unique challenge is the interactional nature of context articulated most prominently by Dourish [59].

−  Building *social competence* into computational agency to ensure that autonomous behaviour complies with the differential rights and privileges at work in the home in order to engender end-user trust. This challenge is distinct from existing work on roles for example in Multi-Agent Systems [60] in that design solutions need to respond to the enacted and fluid ways in which rights and privileges are negotiated on an ongoing basis.

−  Building *transparency* into the data flows that drive autonomous behaviours and data flow controls to further engender user trust. This design challenge can build on initial work in Human-Data Interaction [61], putting forward new models of personal data aligned with GDPR.

−  Building *coordination mechanisms* into autonomous systems to enable users to configure autonomous behaviours around occasional and established patterns of human conduct in the home. While home automation has made significant progress, multi-occupancy is a remaining challenge that has rendered for example a 'learning thermostat' virtually unusable for families [1].

−  Building *control mechanisms* into autonomous systems to enable users to exercise vary levels of control over the behaviour of autonomous systems in the home. While for example occupancy-sensitive home automation has been explored [29, our work seeks to draw attention to the numerous remaining challenges concerning how to best bring the human back into the loop [62].

These are not requirements for autonomous systems in that they do not specify just what should be done or built to address them. Rather, they elaborate problem spaces and topics for design to explore. While we are aware of ongoing research that touches upon the above issues in various ways [e.g., 58, 61, 63, 64], it is not clear, for example, what it would mean in practice to articulate the motives and reasons for autonomous behaviours in the in vivo course of their performance. Would an account have to be provided every time some behaviour occurred or only sometimes and only with respect to certain behaviours? It can be readily anticipated that constant articulation of motives and reasons would become an annoying nuisance having a negative impact on the acceptability of autonomous systems, particularly where relatively trivial behaviours are concerned.

The problem of course is that it is nigh on impossible to say what constitutes 'trivial' (or significant) behaviour in the absence of social context. Thus, a key research challenge here lies not only in building accountability into autonomous behaviours but also in working out how to best support the delivery of accounts to end-users. Ditto building context-awareness, social competence, transparency, coordination and control in autonomous systems, which is to say that what any of these topics might look like and amount to *in practice* has yet to be determined and can only be determined through significant research effort.

Nonetheless, there is sufficient generality built into these design challenges for them to be widely applied in the design of autonomous systems. They might, in effect, be turned into a basic set of design guidelines or fundamental questions such that on any occasion of building an autonomous system, developers might ask themselves if their designs respond to them. For example, in designing an autonomous grocery system its developers might ask:

− Does the system give an account of what motivates its behaviour to the user and the reasons for carrying out particular actions [e.g., that it is ordering XYZ grocery items because you are out of stock and they are the best deal available]?

− Does the system respond to the social circumstances in which it is situated [e.g., presenting accounts of a shopping order at situationally relevant times and places]?

− Does the system display social competence to users in executing its behaviours [e.g., not automatically reordering foodstuffs just because they have run out]?

− Does the system make the data it uses transparent and allow users to control its flow [e.g., not 'sharing' grocery data with large supermarkets and thereby curtailing the flow of adverts and offers?]

− Does the system allow users to coordinate their patterns of behaviour with the system's behaviour [e.g., to 'share' their calendar with the systems as a resource for scheduling reordering?]

− Does the system provide users with granular choices over levels of intelligence and autonomy [e.g., allowing users to delegate certain aspects of grocery shopping to the systems and to retain others for themselves]?

The brackets [] may of course be removed or, perhaps more to the point, their content replaced with specifics concerning the autonomous system to hand. For as with triviality and significance, the question as to what constitutes 'respond to social circumstances', 'display social competence', 'make data use transparent', 'coordinate with patterns of behaviour' and 'provide granular choices' are matters that will need to be worked out with reference to the specificities of an autonomous system and the particular social context into which it is to be situated and used. However, that does not mean the questions cannot be asked, nor answers sought and found.

## 8.5 Limitations
It was suggested in discussion of this paper that the novelty of presenting breaching experiments for design is rather limited; others have beat us to it, as they have with the use of contra-vision. However, the novelty here lies in the intentional configuration of contra-vision scenarios to drive a workshop-based approach to breaching experiments. This contrasts sharply to previous uses of breaching experiments in design to understand 'in the wild' deployments of technology [13, 14, 15, 16], not that previous approaches are homogenous.

Breaching experiments were, to the best of our knowledge, first introduced into design in 2002 by Steve Mann [13], who used wearable computing to create a set of "visible and explicit sousveillance" performances "that follow Harold Garfinkel's ethnomethodological approach to breaching norms." Mann's use of breaching experiments was copybook, i.e., he sought to make trouble and thereby "expose hitherto discreet, implicit, and unquestioned acts of organisational surveillance." The make trouble approach to breaching experiments surfaced again in design in 2009, when Erika Shehan Poole [16] sought to exploit breaching experiments to investigate "existing technology … related practices in domestic settings."

Poole asked participants in her field trial to interact with domestic technology "in ways that potentially disrupted the social norms of the home." More specifically, "each home received weekly 'homework' … intentionally designed to breach household technology installation, usage, and maintenance practices. The assignments the first week served as warm- up to acclimate participants to being in the study … For the second week, the participants were instructed to have the *less* technically oriented adult in the home complete the assignments. This choice was made to disrupt the normal family dynamic …"

The results of Poole's intentional efforts at disruption "provided explanations of *why* problems with technical advice sharing and home technical maintenance persist." Poole subsequently recommends that breaching experiments should be considered as "an asset and an indispensable part of a researcher's toolbox for understanding existing social norms and practices surrounding technology." We do not disagree but would advise caution be exercised when deliberately disrupting the dynamics of any social setting in which the researcher is essentially a guest, and note an alternative approach to breaching experiments in design.

Following Mann, in 2004 Crabtree [14] introduced the notion of breaching experiments not as things that necessarily make trouble and cause disruption, but as an analytic lens on in the wild deployments of novel technology. This approach focuses on explicating through ethnography the "the contingent ways in which novel technology for which no use practice exists is made to work and the interactional practices providing for and organising that work." In a similar vein, in 2008 Tolmie and Crabtree [15] treated novel technological deployments in the home as breaching experiments that "make tacit and taken for granted expectations visible … enabl[ing] us to see how even a simple arrangement of technology can breach ordinary expectations of where technology resides in the home, who owns it, who maintains it, and how user experience of it is accounted for."

However it is not that there are at least two different approaches towards breaching experiments in design or the contrast between them that matters. Despite their differences, previous breaching experiments are oriented, as Poole succinctly puts it, to *existing technology* and related practices, whether it be a novel prototype or technology that has been appropriated at scale and is well-established.

We are not focused on existing technology, whether or not it makes trouble and disrupts or provokes practice by virtue of it having to be 'made to work' in the world. There is no actual system, functionality, connectivity or interactivity in our breaching experiments, only utopian and dystopian envisionings of autonomous systems at work in everyday life. Our breaching experiments are oriented to *future and emerging technology* and the acceptability challenges that confront their adoption in everyday life [34]. Rather than explicate, and even explain existing practice, they instead seek to engage potential end-users in reasoning about the place of future and emerging technologies in their everyday lives and thereby inform us as to key challenges that need to be addressed to make those technologies an acceptable feature of everyday life at an early stage in the design life cycle, before we have built anything. Furthermore, our breaching experiments are done through workshops rather than performances, field trials or ethnography.

It would appear then that there is some novelty to our approach: we are not using breaching experiments to make trouble or disrupt or to provide an analytic lens on existing technology related practice and we are not conducting them in previously practiced ways. The novelty in our approach lies in *re-purposing* tried and tested methods to create utopian and dystopian scenarios that are *designed to disrupt* background expectancies that organise everyday life in familiar settings (in this case the home), and as our findings make perspicuous those expectancies have little to do with existing technology related practice too. The disruption lies in what are essentially incongruous visions of the future depicted by contra-vision scenarios, which in being presented to potential end-users create reality disjunctures that motivate efforts at resolution and repair. It is in the attempt to "resolve incongruities" [9] and repair the reality disjunctures they occasion that ordinarily tacit and unspoken background expectancies are surfaced; expectancies about which people usually have "little or nothing to say" when asked [17] but which this methodological innovation enables early in design.

It is also important to note that it is not the intention in designing breaching experiments that the utopian and dystopian futures depicted in contra-vision scenarios should pre-define problematic aspects of future or emerging technology. Their job, as outlined above, is to disrupt and create a reality disjuncture, whose repair surfaces the taken for granted expectancies that impact future and emerging technologies in everyday life. Does this mean that the contra-visions limit the range of issues brought up by participants, constraining them to the incongruous topics depicted in the contra-vision scenarios? The vignettes presented above would suggest not, insofar as our participants talk and reasoning can be seen to range across a great many matters not depicted in the contra-visions. It would be more apposite, then, to see the contra-visions not as pre-defining design issues or topics but as provocative social objects that elicit multiple background expectancies, which participants themselves come to shape and prioritise in their talk as they go about resolving the incongruities they create.

In breaching taken for granted background expectancies that are usually left unspoken, our utopian and dystopian scenarios have elaborated significant challenges for the design of autonomous systems in the home. It might be countered that these are grand claims to make on the basis of 32 people's say so. However, while articulated by a relatively small number of people, the background expectancies they have expressed do not belong to them alone. As Garfinkel [17] put it,

> *"Almost alone among sociological theorists, the late Alfred Schutz, in a series of classical studies of the constitutive phenomenology of the world of everyday life, described many of these seen but unnoticed background expectancies. He called them the 'attitude of daily life'. He referred to their scenic attributions as the 'world known in common and taken for granted'."*

The attitude of daily life, the world known in common, or in other words what *anyone*, i.e., (with Bittner's caveat [65]) 'any normally competent, wide awake adult' [57] knows about everyday life. There is nothing special about our findings then. They do not speak of and elaborate rarefied knowledge or insight possessed by a privileged few. Rather, the background expectancies articulated by our participants are known in common, shared, used, recognised and relied upon by a much larger cohort and it is for this reason that they elaborate significant challenges for the design of autonomous systems for the home.

## 9. CONCLUSION

People's expectations of domestic autonomous systems are poorly understood at this point in time. We have therefore sought to engage prospective users from the early stages of design, not least because it is widely acknowledged that social expectations concerning the intelligibility and trustworthiness of autonomous systems are as important to their adoption as proposed technological benefits. Our engagement with prospective users exploits a novel design methodology that combines traditional scenario-based design with a contra-vision approach to develop utopian and dystopian visions of the autonomous home that are designed to breach common sense reasoning and surface taken for granted background expectancies that impact the adoption of future and emerging technologies.

This methodological contribution may be exploited more generally but in the case reported here, the forward-looking visions built into our breaching experiments are based on application areas where autonomous systems are expected to make a substantial impact on domestic life. The utopian and dystopian character of these visions disturbs common sense reasoning and creates a reality disjuncture that warrants repair. It is in repairing or resolving the incongruity between utopian and dystopian visions that the usually unspoken background expectancies that people use to make sense of and order familiar settings and scenes in everyday life, including the uses of technology, are articulated and expressed or made visible and thus become available to our consideration.

The substantive contribution of this paper thus reveals that people expect computational and social accountability be built into domestic autonomous systems, along with coordination and control mechanisms. Computational accountability means it is expected that the behaviours of autonomous systems will be made accountable to end-users. Social accountability means it is expected that the behaviours of autonomous systems will not only be legible but situationally appropriate and responsive to social context. Coordination means it is expected that autonomous systems will gear in with discrete patterns of conduct in the home. Control means that it is expected that the intelligence built into autonomous systems will be subject to granular choice.

The results of our breaching experiments do not provide requirements for autonomous systems but acceptability challenges elaborating a number of discrete but interrelated problem spaces and topics that concern the nature of accountability, context-awareness, computational agency, data transparency, and the role of intelligence in the future smart home. What these challenges amount to in practice has yet to be determined. Nonetheless, the expectations are significant and enable fundamental questions to be asked of autonomous systems during their design. Answering them is central to the widespread adoption of autonomous systems in domestic life.

### ACKNOWLEDGMENTS

This work was supported by the Engineering and Physical Sciences Research Council [grant numbers EP/N014243/1, EP/M001636/1].

### REFERENCES

1. Yang R, Newman M (2012) Living with an intelligent thermostat: advanced control for heating and cooling systems. Proceedings of the 2012 Conference on Ubiquitous Computing. ACM, Pittsburg, PA, pp. 1102-1107. https://doi.org/10.1145/2370216.2370449

2. Lehmacher W (2017) The global supply chain: how technology and circular thinking transform our future. Cham, Switzerland, Springer International Publishing. https://doi.org/10.1007/978-3-319-51115-3

3. Stewart B (2016) Promise of a smarter future: why the smart home is catching consumer notice. Dealerscope. https://www.dealerscope.com/post/promise-smarter-future-smart-home-catching-consumer-notice. Accessed 6 February 2019.

4. Smart home market by product: global forecast to 2023. Market and Markets. https://www.marketsandmarkets.com/Market-Reports/smart-homes-and-assisted-living-advanced-technologie-and-global-market-121.html. Accessed 6 February 2019.

5. Yang R, Newman M (2013) Learning from a learning thermostat: lessons for intelligent systems for the home. Proceedings of the 2013 ACM International Joint Conference on Pervasive and Ubiquitous Computing. ACM, Zurich, Switzerland, pp. 93-102. https://doi.org/10.1145/2493432.2493489

6. Rodden T, Fischer JE, Pantidi N, Bachour K, Moran S (2013) At home with agents: exploring attitudes towards future smart energy infrastructure. Proceedings of the SIGCHI Conference on Human Factors in Computing Systems. ACM, Paris, France, pp. 1173-1182. https://doi.org/10.1145/2470654.2466152

7. Wilson C, Hargreaves T, Hauxwell-Baldwin R (2015) Smart homes and their users: a systematic analysis and key challenges. Personal and Ubiquitous Computing, 19(2): 463-476. https://doi.org/10.1007/s00779-014-0813-0

8. Rohracher H (2003) The role of users in the social shaping of environmental technologies. Innovation: The European Journal of Social Science Research, 16(2): 177-192. https://doi.org/10.1080/13511610304516

9. Garfinkel H (1963) A conception of, and experiments with, 'trust' as a condition of stable concerted actions. In: Harvey, O.J (ed.) *Motivation and Social Interaction: Cognitive Determinants*. Ronald Press, New York, pp. 187-238.

10. Garfinkel H (1964) Studies of the routine grounds of everyday activities. Social Problems, 11(3): 225-250. https://doi.org/10.2307/798722

11. Mancini C, Rogers Y, Bandara A, Coe T, Jedrzejczyk L, Joinson A, Price B, Thomas K and Nuseibeh B (2010) Contravision: exploring users' reactions to futuristic technology. Proceedings of the SIGCHI Conference on Human Factors in Computing Systems. ACM, Atlanta, GA, pp. 153-162. https://doi.org/10.1145/1753326.1753350

12. Carroll JM (ed.) (1995) Scenario-based design: envisioning work and technology in system development. New York, Wiley.

13. Mann S, Nolan J, Wellman B (2002) Sousveillance: inventing and using wearable computing devices for data collection in surveillance environments.

Surveillance & Society, 1(3): 331-355. https://doi.org/10.24908/ss.v1i3.3344

14. Crabtree A (2004) Design in the absence of practice: breaching experiments. Proceedings of the 5th ACM Conference on Designing Interactive Systems. ACM, Cambridge, MA, pp. 59-68. https://doi.org/10.1145/1013115.1013125

15. Tolmie P, Crabtree A (2008) Deploying research technology in the home. Proceedings of the 2008 ACM Conference on Computer Supported Cooperative Work. ACM, San Diego, CA, pp. 639-648. https://doi.org/10.1145/1460563.1460662

16. Poole ES (2012) Interacting with infrastructure: a case for breaching experiments in home technology research. Proceedings of the Conference on Computer Supported Cooperative Work. ACM, Seattle, WA, pp. 759-768. https://doi.org/10.1145/2145204.2145319

17. Garfinkel H (1967) Studies in Ethnomethodology. Englewood Cliffs, NJ, Prentice-Hall Inc.

18. Weiser M (1991) The computer for the 21st Century. Scientific American, 265(3): 94-105.

19. Carroll, JM (1997) Scenario-based design. In: Helander MG, Landauer TK, Prabhu PV (eds.) Handbook of Human-Computer Interaction, 2nd edn. Amsterdam, Elsevier, pp. 383-406. ISBN 9780080532882.

20. Carroll, JM (2000) Five reasons for scenario-based design. Interacting with Computers, 13(1): 43-60. https://doi.org/10.1016/S0953-5438(00)00023-0

21. Reeves S (2012) Envisioning ubiquitous computing. Proceedings of the SIGCHI Conference on Human Factors in Computing Systems. ACM, Austin, TX, pp. 1573-1582. https://doi.org/10.1145/2207676.2208278

22. Fischer JE, Crabtree A, Rodden T, Colley J, Costanza E, Jewell M, Ramchurn SD (2016) "Just whack it on until it gets hot" - working with IoT data in the home. Proceedings of the SIGCHI Conference on Human Factors in Computing Systems. ACM, San Jose, CA, pp. 5933-5944. https://doi.org/10.1145/2858036.2858518

23. Ecuity (2013) Smart grids, microgeneration and storage: commercialising the benefits. https://www.ecuity.com/wp-content/uploads/2013/10/SGrid-Report-v_final3.0.pdf. Accessed 6 February 2019.

24. Alper AT, Costanza E, Ramchurn SD, Fischer J, Rodden T, Jennings NR (2016) Tariff agent: interacting with a future smart energy system at home, ACM Transactions on Computer-Human Interaction, 23(4): article no. 25. https://doi.org/10.1145/2943770

25. Seabra JC, Costa MA, Lucena MM (2016) IoT based intelligent system for fault detection and diagnosis in domestic appliances. Proceedings of the 6th International Conference on Consumer Electronics.

IEEE, Berlin, pp. 205-208. https://doi.org/10.1109/ICCE-Berlin.2016.7684756

26. Morollo KH (2015) Smart homes provide convenience and sustainable living. Style. https://www.scmp.com/magazines/style/article/1780091/smart-homes-provide-convenience-and-sustainable-living. Accessed 6 February 2019.

27. Absurdism. Philosophy. http://www.philosophy-index.com/existentialism/absurd.php Accessed 6 February 2019.

28. Stojkoska B, Avramova A, Chatzimisios P (2014) Application of wireless sensor networks for indoor temperature regulation. International Journal of Distributed Sensor Networks. https://doi.org/10.1155/2014/502419

29. Scott J, Bernheim Brush AJ, Krumm J, Meyers B, Hazas M, Hodges S, Villar N (2011) PreHeat: controlling home heating using occupancy prediction. Proceedings of the 13th International Conference on Ubiquitous Computing. ACM, Beijing, China, pp. 281-290. https://doi.org/10.1145/2030112.2030151

30. Hyland L, Crabtree A, Fischer J, Colley J, Fuentes C. (2018) What do you want for dinner? Need anticipation and the design of proactive technologies for the home. Computer Supported Cooperative Work: The Journal of Collaborative Computing and Work Practices, 27 (3-6): 917-46. https://doi.org/10.1007/s10606-018-9314-4

31. Costanza E, Fischer JE, Colley JA, Rodden T, Ramchurn SD, Jennings NR (2014) Doing the laundry with agents: a field trial of a future smart energy system in the home. Proceedings of the SIGCHI Conference on Human Factors in Computing Systems. ACM, Toronto, Canada, pp. 813-822. https://doi.org/10.1145/2556288.2557167

32. Peek S, Aarts S, Wouters E (2017) Can home technology deliver on the promise of independent living? In: van Hoof J, Demiris G, Wouters EJM (eds.) Handbook of Smart Homes, Health Care and Well-Being. Cham, Switzerland, Springer International Publishing, pp. 203-214. https://doi.org/10.1007/978-3-319-01583-5_41

33. Yang R, Newman MW, Forlizzi J (2014) Making sustainability sustainable: challenges in the design of eco-interaction technologies. Proceedings of the SIGCHI Conference on Human Factors in Computing Systems. ACM, Toronto, Canada, pp. 823-832. https://doi.org/10.1145/2556288.2557380

34. Lindley J, Coulton P, Sturdee M (2017) Implications for adoption. Proceedings of the SIGCHI Conference on Human Factors in Computing Systems. ACM, Denver, CO, pp. 265-277. https://doi.org/10.1145/3025453.3025742

35. Pollner M (1975) The very coinage of your brain: the anatomy of reality disjunctures. Philosophy of the Social Sciences, 5(3): 411-430. https://doi.org/10.1177%2F004839317500500304

36. Kitzinger J (1995) Qualitative research: introducing focus groups. British Medical Journal, 311: 299. https://doi.org/10.1136/bmj.311.7000.299

37. Twidale M, Randall D, Bentley R (1994) Situated evaluation for cooperative systems. Proceedings of the 1994 ACM Conference on Computer Supported Cooperative Work. ACM, Chapel Hill, NC, pp. 441-452. https://doi.org/10.1145/192844.193066

38. Braun V, Clarke V (2006) Using thematic analysis in psychology. Qualitative Research in Psychology, 3(2): 77-101. https://doi.org/10.1191/1478088706qp063oa

39. Pollner M (1987) Mundane reason: reality in everyday and sociological discourse. Cambridge, Cambridge University Press. ISBN 9780521321846.

40. Friedman B (1997) Human values and the design of computer technology. Cambridge, Cambridge University Press.

41. Bellotti V, Edwards K (2001) Intelligibility and accountability: human considerations in context-aware systems. Human Computer Interaction, 16(2): 193-212. https://doi.org/10.1207/S15327051HCI16234_05

42. Dourish P, Button G (1998) On 'technomethodology' – foundational relationships between ethnomethodology and system design. Human Computer Interaction, 13(4): 395-432. https://doi.org/10.1207/s15327051hci1304_2

43. Crabtree A, Lodge T, Colley J, Greenhalgh C, Mortier R (2016) Building accountability into the Internet of Things. Social Science Research Network, https://doi.org/10.13140/RG.2.2.27512.44803

44. Shneiderman B. and Plaisant C. (1994) The Future of Graphic User Interfaces: Personal Role Managers, In Bcs hci, pp. 3-8. https://doi.org/10.1017/CBO9780511600821.002

45. Porcheron M., Fischer J., Reeves S., and Sharples S. (2018) Voice Interfaces in Everyday Life. In Proceedings of the 2018 CHI Conference on Human Factors in Computing Systems, p. 640. https://doi.org/10.1145/3173574.3174214

46. Bowden, K.K., Nilsson, T., Spencer, C.P., Cengiz, K., Ghitulescu, A. and van Waterschoot, J.B., 2017. I Probe, Therefore I Am: Designing a Virtual Journalist with Human Emotions. arXiv preprint arXiv:1705.06694.

47. Dautenhahn, K., Nehaniv, C. L., Walters, M. L., Robins, B., Kose-Bagci, H., Mirza, N. A., & Blow, M. (2009). KASPAR–a minimally expressive humanoid robot for human–robot interaction research. Applied Bionics and Biomechanics, 6(3-4), 369-397. https://doi.org/10.1080/11762320903123567

48. Crabtree A, Lodge T, Colley J, Greenhalgh C, Glover K, Haddadi H, Amar Y, Mortier R, Li Q, Moore J, Wang L, Yada P, Zhao J, Brown A, Urquhart L, McAuley D (2018) Building accountability into the Internet of Things: the IoT Databox model. Journal of Reliable Intelligent Environments, 4(1): 39-55. https://doi.org/10.1007/s40860-018-0054-5

49. Stumpf S, Burnett M, Pipek V, Wong WK (2012) End-user interactions with intelligent and autonomous systems. Extended Abstracts on Human Factors in Computing Systems. ACM, Austin, TX, pp. 2755-2758. https://doi.org/10.1145/2212776.2212713

50. Lim B, Dey A, Avrahami D (2009) Why and why not explanations improve the intelligibility of context-aware intelligent systems. Proceedings of the SIGCHI Conference on Human Factors in Computing Systems. ACM, Boston, MA, pp. 2119-2128. https://doi.org/10.1145/1518701.1519023

51. Davis FD (1985) A technology acceptance model for empirically testing new end-user information systems: theory and results. Dissertation, Massachusetts Institute of Technology. https://dspace.mit.edu/handle/1721.1/15192. Accessed 6 February 2019.

52. Davis FD, Bagozzi RP, Warshaw PR (1989) User acceptance of computer technology: a comparison of two theoretical models. Management Science, 35(8): 982-1003. https://doi.org/10.1287/mnsc.35.8.982

53. Hornbæk K, Hertzum M (2017) Technology acceptance and user experience: a review of the experiential component in HCI. ACM Transactions on Computer-Human Interaction, 24(5): article no. 33. https://doi.org/10.1145/3127358

54. Hassenzahl M, Tractinsky N (2006) User experience: a research agenda. Behaviour and Information Technology, 25(2): 91-97. https://doi.org/10.1080/01449290500330331

55. Kim HC (2015) Acceptability engineering: the study of user acceptance of innovative technologies. Journal of Applied Research and Technology, 13(2): 230-237. https://doi.org/10.1016/j.jart.2015.06.001

56. Pitts M (2016) We need trust to innovate. Gov.UK: Innovate UK. https://innovateuk.blog.gov.uk/2016/08/17/we-need-trust-to-innovate. Accessed 6 February 2019.

57. Whitworth B (2005) Polite computing. Behaviour and Information Technology, 24(5): 353-363. https://doi.org/10.1080/01449290512331333700

58. Suchman L, Weber J (2016) Human machine autonomies. In: Bhuta N, Beck S, Geiβ R, Liu H, Kreβ C (eds.) Autonomous weapons systems: law, ethics,

policy. Cambridge, Cambridge University Press, pp. 75–102. https://doi.org/10.1017/CBO9781316597873.004

59. Dourish P (2004) What we talk about when we talk about context. Personal and Ubiquitous Computing 8(1): 19-30. https://doi.org/10.1007/s00779-003-0253-8

60. Ramchurn SD, Huynh D, Jennings NR (2004) Trust in multi-agent systems. The Knowledge Engineering Review, 19(1): 1-25. https://doi.org/10.1017/S0269888904000116

61. Mortier R, Haddadi H, Henderson T, McAuley D, Crowcroft J, Crabtree A (2016) Human Data Interaction. In: The Encyclopedia of Human-Computer Interaction. Denmark, Interaction Design Foundation, Chapter 41.

62. Fischer JE, Greenhalgh C, Jiang W, Ramchurn SD, Wu F, Rodden T (2017) In-the-loop or on-the-loop? Interactional arrangements to support team coordination with a planning agent. Concurrency and Computation: Practice and Experience. https://doi.org/10.1002/cpe.4082

63. Dourish P, Bell G (2014) Resistance is futile: reading science fiction alongside ubiquitous computing. Personal and Ubiquitous Computing, 18(4): 69-778. https://doi.org/10.1007/s00779-013-0678-7

64. Crabtree A, Lodge T, Colley J, Greenhalgh C, Mortier R, Haddadi H (2016) Enabling the new economic actor: data protection, the digital economy, and the Databox. Personal and Ubiquitous Computing, 20(6): 947-957. https://doi.org/10.1007/s00779-016-0939-3

65. Bittner E (1973) Objectivity and realism in sociology. In: Psathas G (ed.) Phenomenological sociology. Ann Arbor, MI, Wiley, pp. 109-125.